\documentclass[prl,twocolumn,showkeys,showpacs,preprintnumbers,amsmath,amssymb]{revtex4}

\usepackage{graphicx}
\usepackage{dcolumn}
\usepackage{bm}
\usepackage{times}

\def\stamp{--- {\bf \today} --- {\bf \jobname.tex}}

\def\bea{\begin{eqnarray}}
\def\eea{\end{eqnarray}}
\def\beq{\begin{equation}}
\def\eeq{\end{equation}}

\def\BE{\begin{equation}}
\def\EE{\end{equation}}
\def\spa#1.#2{\left\langle#1\,#2\right\rangle}
\def\spb#1.#2{\left[#1\,#2\right]}
\def\spba#1.#2.#3{\left[#1|#2|#3\right\rangle}
\def\spab#1.#2.#3{\left\langle#1|#2|#3\right]}
\def\spaa#1.#2.#3{\left\langle#1|#2|#3\right\rangle}
\def\spbb#1.#2.#3{\left[#1|#2|#3\right]}
\def\lor#1.#2{\left(#1\,#2\right)}

\def\Year{\expandafter\eatPrefix\the\year}
\newcount\hours \newcount\minutes
\def\monthname{\ifcase\month\or
January\or February\or March\or April\or May\or June\or July\or
August\or September\or October\or November\or December\fi}
\def\shortmonthname{\ifcase\month\orx
Jan\or Feb\or Mar\or Apr\or May\or Jun\or Jul\or
Aug\or Sep\or Oct\or Nov\or Dec\fi}

\def\TimeStamp{\hours\the\time\divide\hours by60%
\minutes -\the\time\divide\minutes by60\multiply\minutes by60%
\advance\minutes by\the\time%
${\rm \shortmonthname}\cdot   \if\day<10{}0\fi\the\day\cdot   \the\year
\qquad\the\hours:\if\minutes<10{}0\fi\the\minutes$}

\newskip\humongous \humongous=0pt plus 1000pt minus 100pt

\newif\ifdtup

\newcounter{eqnumber}[section]

\newbox\charbox
\newbox\slabox

\def\spa#1.#2{\left\langle#1\,#2\right\rangle}
\def\spb#1.#2{\left[#1\,#2\right]}
\def\lor#1.#2{\left(#1\,#2\right)}

\catcode`@=11  

\def\lsl{\not{\hbox{\kern-2.3pt $\ell$}}}
\def\ksl{\not{\hbox{\kern-2.3pt $k$}}}

\def\spa#1.#2{\left\langle#1\,#2\right\rangle}
\def\spb#1.#2{\left[#1\,#2\right]}
\def\lor#1.#2{\left(#1\,#2\right)}

\def\sand#1.#2.#3{%
  \left\langle\smash{#1}{\vphantom1}\right|{#2}%
  \left|\smash{#3}{\vphantom1}\right\rangle}
\def\sandp#1.#2.#3{%
  \left\langle\smash{#1}{\vphantom1}^{-}\right|{#2}%
  \left|\smash{#3}{\vphantom1}^{+}\right\rangle}
\def\sandpp#1.#2.#3{%
  \left\langle\smash{#1}{\vphantom1}^{+}\right|{#2}%
  \left|\smash{#3}{\vphantom1}^{+}\right\rangle}
\def\sandmm#1.#2.#3{%
  \left\langle\smash{#1}{\vphantom1}^{-}\right|{#2}%
  \left|\smash{#3}{\vphantom1}^{-}\right\rangle}
\def\sandpm#1.#2.#3{%
  \left\langle\smash{#1}{\vphantom1}^{+}\right|{#2}%
  \left|\smash{#3}{\vphantom1}^{-}\right\rangle}
\def\sandmp#1.#2.#3{%
  \left\langle\smash{#1}{\vphantom1}^{-}\right|{#2}%
  \left|\smash{#3}{\vphantom1}^{+}\right\rangle}

\begin{document}

\title{Gravity and Yang-Mills Amplitude Relations}

\author{N. E. J. Bjerrum-Bohr$^a$, Poul H. Damgaard$^a$, Bo Feng$^b$ and Thomas S{\o}ndergaard$^a$}
\affiliation{$^a$Niels Bohr International Academy and Discovery
Center,\\ The Niels Bohr Institute, Blegdamsvej 17, DK-2100,
Copenhagen \O, Denmark}\email{bjbohr;phdamg;tsonderg@nbi.dk}
\affiliation{\\
$^b$Center of Mathematical Science, Zhejiang University,
Hangzhou, China.}\email{b.feng@cms.zju.edu.cn}


\date{\today}

\begin{abstract}
Using only general features of the S-matrix and
quantum field theory, we prove by induction the
Kawai-Lewellen-Tye relations that link products of gauge
theory amplitudes to gravity amplitudes at tree level.
As a bonus of our
analysis, we provide a novel and more symmetric form
of these relations. We also establish an infinite tower of
new identities between amplitudes in gauge theories.
\end{abstract}

\maketitle

{\em Introduction.}~ A most astonishing connection between
Einstein gravity and color-ordered Yang-Mills tree amplitudes is
provided by the Kawai-Lewellen-Tye (KLT) relations~\cite{KLT}.
The infinite sequence of these KLT-relations was discovered as
a consequence of factorizing a closed-string amplitude into a
product of open-string amplitudes and subsequently taking the
field theory limit. For a nice introduction to the details of
this, see, {\it e.g.}, ref.~\cite{Bern:2002kj} and references
therein. At the quantum field theory level the KLT-relations
present a mysterious puzzle, since neither the
Einstein-Hilbert nor the Yang-Mills Lagrangians provide any
hints at their origin. In fact, the two theories appear to be
very dissimilar in structure. The gravity Lagrangian yields
perturbative Feynman rules with an infinite series
of higher-point graviton vertices while the Yang-Mills
Lagrangian terminates at four-point vertices. The gravity
Lagrangian has general coordinate covariance, while Yang-Mills
theory has local gauge invariance. It could
seem that something close to a miracle would be required to relate
the two associated S-matrices. It is one of
the great achievements of string theory that it
inspires a re-organization of the perturbative expansions
that sheds completely new light on this. In a related
development, it has been shown in
ref.~\cite{BjerrumBohr:2009rd} how string theory can be used to
derive the conjectured Bern-Carrasco-Johansson (BCJ) identities
in gauge theories with and without
matter~\cite{Bern:2008qj,Sondergaard:2009za}, see also~\cite{Stie}. While this
illustrates again the power of string-based techniques, it also
highlights the need for a similar understanding directly at the
field theory level. Very recently, the BCJ-relations were
proven~\cite{Feng:2010my} using only quantum field theory,
based on the method of on-shell
recursion~\cite{Britto:2004ap,Britto:2005fq}. There has also been
attempts at more conventional ways to understand the KLT-relations
at the Lagrangian level~\cite{Bern:1999ji} through judicious
choices of gauges. One possibility is a reformulation of the
Yang-Mills Lagrangian through the addition of spurious
vertices up to infinite order~\cite{Bern:2010yg}. An alternative
path consists in writing the gauge theory amplitudes explicitly
in terms of a selected set of pole structures. A squaring relation
between gravity and gauge theory poles, conjectured to hold
to all orders in ref.~\cite{Bern:2008qj}, can then be
proven~\cite{Bern:2010yg}.

In the light of recent progress, we will here take a fresh
approach to the KLT-relations. We prove these relations
using only quantum field theory and general properties of the
S-matrix~\cite{Benincasa:2007xk}. As
a spin-off, we uncover a new series of highly non-trivial
relations entirely on the gauge theory side. These identities
are non-linear and involve products of different helicity
configurations of gauge theory amplitudes. We provide a novel
form of the KLT-relations as well. It has a higher degree
of manifest symmetry than the one previously suggested in the
literature~\cite{Bern:1998ug}. Interestingly, the two different
forms are precisely related to each other via the
BCJ-relations.

\medskip
{\em Gravity from Gauge Theory and New Relations between Gauge
Theory Amplitudes.}~ We denote an
$n$-point gravity amplitude of fixed helicity by $M_n(1,2,\ldots,n)$ and let
$A_n(1,2,\ldots,n)$ and $\widetilde{A}_n(1,2,\ldots,n)$ stand for $n$-point
color-ordered gauge theory amplitudes of fixed helicity.
Both classes of amplitudes have been
stripped of coupling constants since it is trivial to reinstate them.
We also denote, as usual, $s_{12\ldots i}
\equiv (p_1 + \ldots + p_{i})^2$.
Our two main results are\smallskip
\begin{equation}
\begin{split}
M_n(1,2,\ldots,n)=&\\
&\hspace{-2.5cm}(-1)^n\!\!\sum_{\gamma,\beta}{\widetilde{A}_n(n,\gamma_{2,n\!-\!1},1)
{\cal S}[
\gamma_{2,n\!-\!1}|\beta_{2,n\!-\!1}]
A_n(1,\beta_{2,n\!-\!1},n)\over s_{123\ldots
(n\!-\!1)}}\label{newKLT}\,,
\end{split}\end{equation}
\begin{equation}
\begin{split}
0\!=\!\sum_{\gamma,\beta}\!{\widetilde{A}_n(n,\gamma_{2,j^+\!,n\!-\!1},1)
{\cal S}[ \gamma_{2,n\!-\!1}|\beta_{2,n\!-\!1}]
A_n(1,\beta_{2,j^-\!,n\!-\!1},n)\over s_{123\ldots
(n\!-\!1)}}\label{vanishing}\,,
\end{split}
\end{equation}%
both of which will be proven by induction. We have chosen one
arbitrary external leg $j$ to have opposite helicity in
eq.~(\ref{vanishing}). The ordering of legs $2,3,\ldots,n-1$ in
the amplitude ${\widetilde{A}_n}$ is denoted $\gamma_{2,n-1}$
and $\gamma_{2,j^\pm,n-1}$, where $j^\pm$ indicates that leg
$j$ has been assigned a specific helicity. We note here since we will use it
later that it is also possible that the assigned helicity leg $j^\pm$
is either leg 1 or $n$. The corresponding ordering in the amplitude $A_n$
is denoted by $\beta_{2,n-1}$ and $\beta_{2,j^\pm,n-1}$ and we
sum over all permutations of both $\gamma$ and $\beta$. The
function ${\cal S}$ is defined by
\begin{equation}
{\cal S}[i_1,\ldots,i_k|j_1,\ldots,j_k]\! \equiv\!\!
\prod_{t=1}^k \big(s_{i_t 1}\!+\!\sum_{q>t}^k \theta(i_t,i_q)
s_{i_t i_q}\big) \label{Sdef}\,,
\end{equation}
where $\theta(i_a,i_b)$ is 0 if $i_a$ sequentially comes before
$i_b$ in
$\{j_1,\ldots,j_k\}$, and otherwise it is 1. To illustrate,
${\cal S}[2|2] = s_{12}, {\cal S}[23|23] = s_{12}s_{13},
{\cal S}[23|32] = s_{13}(s_{12} + s_{23})$, and so on.
\medskip

The function $\cal S$ has some nice properties that all
follow from its definition
(\ref{Sdef}) by use of elementary algebra.
These properties will play an essential role in what
follows. In particular,
\begin{equation}
{\cal S}[i_1,\ldots,i_k|j_1,\ldots,j_k]  = {\cal
S}[j_k,\ldots,j_1|i_k,\ldots,i_1]\,,\label{Ssym}
\end{equation}
which ensures that the expressions (\ref{newKLT}) and (\ref{vanishing})
are completely symmetric in $\widetilde{A}_n$ and $A_n$.
It is also convenient to introduce an auxiliary function,
\begin{equation}
 \mathcal{S}_{P}[i_1,\ldots,i_k|j_1,\ldots,j_k] =
\prod_{t=1}^k \big( s_{i_tP} +
\sum_{q>t}^k\theta(i_t,i_q)s_{i_ti_q}\big)\,, \label{SPdef}
\end{equation}
which coincides with ${\mathcal S}$ except for the fact that
the momentum of leg 1 has been replaced by a sum of momenta, $P \equiv
p_1+p_2+\ldots +p_m$ with $P^2=0$,
which not necessarily involves any of the momenta in the brackets.
We point out that one has the factorization
\begin{equation}
\mathcal{S}[\gamma_{q\!+\!1,k}\sigma_{2,q}|
\alpha_{2,q}\beta_{q\!+\!1,k}] =
\mathcal{S}[\sigma_{2,q}|\alpha_{2,q}]\! \times\!
\mathcal{S}_{P}[\gamma_{q\!+\!1,k}|\beta_{q\!+\!1,k}]\,,\label{cuteSrel}
\end{equation}
with $P= p_1+p_2+\ldots +p_q$.
\medskip

\noindent We now note the following:
\begin{itemize}
\item Eq.~(\ref{newKLT}) provides the general $n$-point result
    for the field theory limit of the
    KLT-relations~\cite{KLT}.
\item Eq.~(\ref{vanishing}) provides a new set of identities
    between gauge theory amplitudes of different helicity
    configurations.
\end{itemize}

An unusual property of the expressions (\ref{newKLT}) and
(\ref{vanishing}) is that they appear to be singular on-shell.
However, the singularity due to
$s_{12\ldots (n-1)}$ is only apparent: It is always cancelled
by a similar factor in the numerator. This will be explained below.
\medskip

A different form of the KLT-relations was conjectured in
ref.~\cite{Bern:1998ug}. Our new expression (\ref{newKLT}), which
keeps only two legs fixed while summing over all permutations of the remaining legs, is
more symmetric and therefore more convenient for our purpose.
\medskip

{\em Proof by induction:}~ We will treat the cases (\ref{newKLT}) and
(\ref{vanishing}) in parallel.
To handle the apparent singularity of
$s_{12\ldots (n-1)}$ we need to regularize both expressions
(\ref{newKLT}) and (\ref{vanishing}). We could choose the following
regularization:
\begin{align}
p_1 &\rightarrow p_1 - x q\,,\cr
p_n & \rightarrow  p_n + x q\,,
\label{regu}
\end{align}
with a parameter $x$, $p_1\cdot q=0$ and $q^2=0$, but $q\cdot p_n\neq 0$.
This keeps $p_1^2=0$, respects overall momentum conservation,
but makes $p_n^2=s_{12\ldots (n-1)}\neq 0$. We recover the physical
amplitudes in the limit $x \to 0$.\medskip\medskip

Before proceeding further, we make a few more remarks regarding the regularization
(\ref{regu}) and how one cancels the pole $s_{12\ldots(n-1)}$.
Interestingly,
the numerators of (\ref{newKLT}) and (\ref{vanishing}) vanish on-shell
precisely because of BCJ-relations. In detail, for each $\gamma_{2,n\!-\!1}$ permutation,
\begin{equation}
\sum_{\beta}{\cal S}[\gamma_{2,n\!-\!1}|\beta_{2,n\!-\!1}]
A_n(1,\beta_{2,n\!-\!1},n) = 0\,,
\end{equation}
{\em is} in general a combination of BCJ-relations. One can write an analogous
relation for $\widetilde{A}_n$ by means of a $\gamma$-permutation sum.
Once the full numerators in (\ref{newKLT})
and (\ref{vanishing}) are regularized
according to, for instance, eq. (\ref{regu}), they do not vanish. Lifting the
regularization from terms that remain finite in the $x \to 0$ limit, one
can systematically exploit on-shell BCJ-relations to factor out the
needed factor of $p_n^2 = s_{12\ldots(n-1)}$ which cancels the would-be pole.
Afterwards
the limit can safely be taken in all remaining terms. This reduction, however,
destroys the larger manifest permutation symmetry of
(\ref{newKLT}) and (\ref{vanishing}), and
the reduced expresssion is therefore not the most convenient form for
a BCFW-analysis \cite{Britto:2004ap,Britto:2005fq}.
We have checked up to $n=8$ that the reduced expression
agrees with the general formula suggested in ref. \cite{Bern:1998ug}.
More details on this will be presented elsewhere
\cite{paper}.\medskip

When $n=3$ both eq.~(\ref{newKLT}) and
eq.~(\ref{vanishing}) hold trivially. The
right-hand side of both equations becomes,
after removing the regularization, $-\widetilde{A}_3(3,2,1)A_3(1,2,3)$.
On-shell, for real momenta, these 3-point amplitudes
vanish, and both identities are satisfied. For on-shell complex momenta
the relation (\ref{newKLT}) reads $M_3(1,2,3) =
-\widetilde{A}_3(3,2,1)A_3(1,2,3)$, which is indeed the
correct three-graviton amplitude
\cite{Benincasa:2007xk} (both sides vanish when all helicities are equal).
For $n=4$ the right-hand side becomes
\begin{align}
s_{12}&\widetilde{A}_4(4,3,2,1)
\!\frac{ (s_{13}\!+\!s_{23})A_4(1,2,3,4)
\!+\!s_{13}A_4(1,3,2,4)}{s_{123}}+ \nonumber \\
s_{13}&\widetilde{A}_4(4,2,3,1)
\!\frac{(s_{12}\!+\!s_{23})
\!A_4(1,3,2,4)
\!+\!s_{12} A_4(1,2,3,4)}{s_{123}}\,,
\end{align}
where we have collected pieces so that the mentioned structure of
BCJ-relations appears in the numerator.
We can then take the limit $x\rightarrow 0$ in the two terms
$s_{12}\widetilde{A}(4,3,2,1)$ and $s_{13}\widetilde{A}(4,2,3,1)$ separately,
use the on-shell
BCJ-relation $s_{12}\widetilde{A}(4,3,2,1)=s_{13}\widetilde{A}(4,2,3,1)$
and collect terms to get an overall factor of $s_{123}$ which
precisely cancels the denominator. The regularization can then be
removed. Doing these steps we are left with
$$
s_{12}\widetilde{A}_4(4,3,2,1) [A_4(1,2,3,4)+A_4(1,3,2,4)]\,,
$$
which by use of standard amplitude relations can be written as the
more familiar KLT expression $-s_{12}\widetilde{A}_4(1,2,3,4)A_4(1,2,4,3)$.
\medskip

For the identities to be of interest, we of course
take helicities so that the amplitudes are non-vanishing
to begin with.
Flipping the helicity of one of the external legs in either $\widetilde{A}_4$
or $A_4$ will cause those amplitudes to vanish, and eq. (\ref{vanishing})
is thus trivially satisfied. If we do
not flip the helicity of one of the legs, we see by explicit
computation that we get
the four-graviton amplitude $M_4(1,2,3,4)$ for the chosen helicities.

The origin of the cancellation of the $s_{12\ldots(n-1)}$-pole
hinges on the basis of amplitudes being of size $(n-3)!$
\cite{BjerrumBohr:2009rd}, while the permutation sums in
(\ref{newKLT}) and (\ref{vanishing}) keep only {\em two} legs,
1 and $n$, fixed. The sums are therefore overcomplete and
redundant.
\medskip

After these preliminary remarks, we are now ready to prove the general
relations by induction. We have already demonstrated by explicit
computations that the relations hold for both real and complex momenta
when $n=3$ and $n=4$.
We next
assume that eq.~(\ref{newKLT}) and eq.~(\ref{vanishing}) both hold for
$n-1$. Doing a BCFW-shift in
legs $1$ and $n$, we consider the following contour integral
\begin{align} 0 =
\oint \frac{dz}{z}M_n(z) = M_n(0) +
(\mathrm{residues\:\:for}\:\:z\neq 0)\,.
\end{align}
If there should be boundary terms to the integral, they are ignored
here. It is known that the $n$-point gravity amplitudes $M_n$ have
sufficiently rapid fall-off at infinity to exclude boundary
terms~\cite{Benincasa:2007qj,zscaling,ArkaniHamed:2008gz}.

For the $z\neq 0$ residues, we consider separately the following
two classes of contributions:\medskip
\begin{itemize}
\item[(A)]
The pole appears in only one of the amplitudes $\widetilde{A}_n$ and $A_n$.
\item[(B)] The pole appears in both
amplitudes  $\widetilde{A}_n$ and $A_n$.
\end{itemize}\medskip
We start with the case (A). By symmetry, we need to consider only
$\widetilde{A}_n$ having the pole. When the pole instead sits in
$A_n$ the reasoning is identical.
For ease of notation we will omit the explicit writing of
$\lim_{x\rightarrow 0}$ in what follows.

Considering eq.~(\ref{newKLT}), the residue of the pole
$s_{\widehat{1}2..k}$ can be calculated from
$-\lim_{z\rightarrow z_{12..k}} \big[ s_{\widehat{1}2..k}(z)
M_n(z) \big]/s_{12..k}$, where $z_{12..k}$ is the $z$-value
that makes $s_{\widehat{1}2..k}$ go on-shell. We hence get
\begin{align}\label{partA}
&\nonumber\frac{(-1)^{n+1}}{s_{\widehat{1}2..n-1}} \sum_{\gamma,\sigma,\beta}
\frac{\sum_h\widetilde{A}_{n-k+1}(\widehat{n},\gamma,-\widehat{P}^h)
\widetilde{A}_{k+1}(\widehat{P}^{-h},\sigma,\widehat{1})}{s_{12..k}}
\\ & \hspace{1cm}\quad\times
\mathcal{S}[\gamma\sigma|\beta_{2,n-1}]
A_n(\widehat{1},\beta_{2,n-1},\widehat{n})\,,
\end{align}
where we have introduced the short-hand notation $\gamma \equiv
\gamma_{k+1,n-1}$ and $\sigma \equiv \sigma_{2,k}$. Now using
a factorization analogous to eq.~(\ref{cuteSrel}) we can rewrite
\begin{align}
\mathcal{S}[\gamma\sigma|\beta_{2,n-1}] =
\mathcal{S}[\sigma|\rho_{2,k}]\! \times\! \text{(a factor
independent of } \sigma)\,,
\end{align}
where $\rho_{2,k}$ denotes the relative
ordering of legs $2,3,\ldots,k$ in
$\beta$. We thus see that eq.~\eqref{partA} contains
\begin{align}
\sum_{\sigma}\widetilde{A}_{k+1}(\widehat{P}^{-h},\sigma,\widehat{1})
\mathcal{S}[\sigma|\rho_{2,k}] = 0\,,
\end{align}
which is zero at $z=z_{12..k}$, as indicated.
It is important for this argument that $A$ does
not have a pole at $s_{12..k}$ since such a pole could cancel
the above zero. We hereby conclude that all terms coming from the class (A)
above will not
contribute to the residues. Going through the analogous argument
for eq.~(\ref{vanishing}) we conclude similarly about the
case (A) there.

We now turn to the class (B). Again we consider first
eq.~(\ref{newKLT}) and the $s_{12..k}$ pole contribution. Here
both $\widetilde{A}$ and $A$ have the pole. Similar to
the short-hand notation of $\gamma$ and $\sigma$ above, we will
also introduce $\beta \equiv \beta_{k+1,n-1}$ and $\alpha
\equiv \alpha_{2,k}$.

When both amplitudes have the pole, the residue takes the
following form (we have in the two equations below
suppressed the subscript index on $A$ and $\widetilde{A}$
to avoid unnecessary cluttering of the expressions)
\begin{widetext}
\begin{align}
\frac{(-1)^{n+1}}{s_{\widehat{1}2...(n-1)}}
\sum_{\gamma,\beta,\sigma,\alpha}
\left[ \frac{\sum_h \widetilde{A}(\widehat{n},\gamma,\widehat{P}^{-h})
\widetilde{A}(-\widehat{P}^{h},\sigma,\widehat{1})}{s_{12..k}} \right]
 \mathcal{S}[\gamma\sigma|\alpha\beta]
\left[ \frac{\sum_h A(\widehat{1},\alpha,-\widehat{P}^h)
A(\widehat{P}^{-h},\beta,\widehat{n})}
{s_{\widehat{1}2..k}} \right]\,,
\end{align}
where one of the shifted $s_{\widehat{1}2..k}$ poles have been
replaced by an unshifted pole $s_{12..k}$ from calculating the
single-pole residues. We now wish to collect pieces so that
lower-point amplitude combinations $\widetilde{A}_kA_k$ appear
in forms ready for a BCFW-interpretation at these lower points.
Noting that $s_{\widehat{1}2..n-1} =
s_{\widehat{P}k+1..n-1}$, and using
$\mathcal{S}[\gamma\sigma|\alpha\beta]
=\mathcal{S}[\sigma|\alpha] \times
\mathcal{S}_{\widehat{P}}[\gamma|\beta]$, this can be achieved
by writing the above contribution as
\begin{align} &
\frac{(-1)^{n+1}}{s_{12..k}} \sum_h\left[ \Bigg(\sum_{\sigma,\alpha}
\frac{\widetilde{A}(-\widehat{P}^{h},\sigma,\widehat{1})\mathcal{S}
[\sigma|\alpha]A(\widehat{1},\alpha,-\widehat{P}^h)}
{s_{\widehat{1}2..k}} \Bigg)
\Bigg(\sum_{\gamma,\beta}
\frac{\widetilde{A}(\widehat{n},\gamma,\widehat{P}^{-h})
\mathcal{S}_{\widehat{P}}[\gamma|\beta]
A(\widehat{P}^{-h},\beta,\widehat{n})}{s_{\widehat{P}k+1..(n-1)}}
\Bigg) \right] + (h,-h)\,, \label{partB}
\end{align}
where $(h,-h)$ means the same expression again, but with the
$(-\widehat{P}^h,-\widehat{P}^h)$ in the first parenthesis and
$(\widehat{P}^{-h},\widehat{P}^{-h})$ in
the second parenthesis replaced by $(-\widehat{P}^h,-\widehat{P}^{-h})$ and
$(\widehat{P}^{h},\widehat{P}^{-h})$, respectively. These are
\textit{mixed-helicity} terms. The appearance of mixed-helicity
terms is what prevents an immediate recombination into lower-point
$M_n$-amplitudes.
\\
\end{widetext}

Fortunately, the first term of eq.~\eqref{partB} is nothing but the product
of two lower-point expressions of eq.~\eqref{newKLT}. We remind the reader
that we have suppressed the overall $\lim_{x\to 0}$ and $\lim_{z\to z_{12..k}}$
at all steps in the derivation, ensuring that both expressions are well-defined,
\textit{i.e.}
\begin{align}
-\frac{\sum_h M_{k+1}(\widehat{1},2,\ldots,k,-\widehat{P}^h)
M_{n-k+1}(\widehat{P}^{-h},k+1,\ldots,\widehat{n})}{s_{12..k}}\,.
\end{align}
Summing over all the permutations, these pieces precisely build up
the amplitude $M_n$ by means of on-shell recursion.
The second term of eq.~\eqref{partB} is a mixed-helicity expression, identical
to the type of relations given by eq.~(\ref{vanishing}).

The proof of eq.~(\ref{vanishing}) for the mixed-helicity relations
follows exactly the same steps as in the derivation of (\ref{newKLT})
above. The only
difference is
that in eq.~\eqref{partB} each helicity sum has a part
that includes a lower-point mixed-helicity relation. By our
induction hypothesis, this is zero.
Because of the manifest permutation symmetry in all legs,
except for the shifted legs 1 and $n$,
every other contribution to the residue follows from this case by a
permutation. The mixed-helicity terms in eq.~(\ref{partB})
hence vanish. This therefore concludes our
proof by induction of both our new form of the KLT-relations eq.~(\ref{newKLT})
and the new relations between gauge theory amplitudes
eq.~(\ref{vanishing}).
\medskip

{\em Conclusions.}~ We have discovered a new and more symmetric
form of the KLT-relations which relate tree-level gravity
amplitudes to products of gauge theory amplitudes. In the process
we have
uncovered a series of non-linear identities among gauge
theory amplitudes where helicities are flipped. We have proven by induction
both sets of relations using on-shell recursion methods.
Our proof does
not rely on any other properties of the amplitudes than those provided
by quantum field theory and general assumptions about the
S-matrix.

We have here concentrated only on the basic identity between pure
Yang-Mills theory and gravity because this is the perhaps most startling
result. As will be discussed in detail elsewhere \cite{paper},
one can straightforwardly
extend the analysis to include all amplitudes from full
supersymmetric multiplets on the gauge theory and gravity sides,
{\it i.e.} ${\cal N}=4$ super Yang-Mills theory and ${\cal N}=8$
supergravity, respectively. This includes
analogous relations for amplitudes
of mixed particle content in
$A_n$ and $\widetilde{A}_n$ \cite{Bern:1999bx}.
The new series of gauge theory identities (\ref{vanishing})
is only a particular example of a more general series of
identities where also more than one pair of helicities can be flipped
\cite{paper}. Although these new identities
have natural interpretations in a KLT-like language, they
are nevertheless on a different footing. It would be nice to
also understand these new identities in the light of string theory.

In a broader perspective, it should be of interest
to understand the significance of the relation between
gravity amplitudes and gauge theory amplitudes at loop
level as well. There has very recently been interesting progress
in this direction \cite{Bern:2010ue}. Also here the method of
on-shell recursion may provide new insight.

{\sc Acknowledgments:}~(BF) would like to acknowledge funding
from Qiu-Shi, the Fundamental Research Funds for the Central Universities,
as well as Chinese NSF funding under contract
No.10875104.

\end{document}